\documentclass[journal,onecolumn]{IEEEtran}

\usepackage{graphicx}
\usepackage{empheq}
\usepackage{amsmath}
\usepackage{amsfonts}
\usepackage{amssymb}
\usepackage{caption}
\usepackage{mathtools}
\DeclareMathOperator*{\argmax}{argmax}

\usepackage[ruled,vlined]{algorithm2e}
\usepackage[noend]{algpseudocode}
\usepackage{romannum}
\usepackage{color}
\newcommand{\etal}{\textit{et al}.}

\title{Delay-aware and Energy-Efficient Computation Offloading in Mobile Edge Computing Using Deep Reinforcement Learning}

\author {Laha Ale, Ning Zhang,
         Xiaojie~Fang,
         Xianfu~Chen,
         Shaohua~Wu,
         Longzhuang Li

}

\begin{document}

\maketitle

\begin{abstract}
Internet of Things (IoT) is considered as the enabling platform for a variety of promising applications, such as smart transportation and smart city, where massive devices are interconnected for data collection and processing. These IoT applications pose a high demand on storage and computing capacity, while the IoT devices are usually resource constrained. As a potential solution, mobile edge computing (MEC) deploys cloud resources in the proximity of IoT devices so that their requests can be better served locally. In this work, we investigate computation offloading in a dynamic MEC system with multiple edge servers, where computational tasks with various requirements are dynamically generated by IoT devices and offloaded to MEC servers in a time-varying operating environment (e.g., channel condition changes over time). The objective of this work is to maximize the completed tasks before their respective deadlines and minimize energy consumption. To this end, we propose an end-to-end Deep Reinforcement Learning (DRL) approach to select the best edge server for offloading and allocate the optimal computational resource such that the expected long-term utility is maximized. The simulation results are provided to demonstrate that the proposed approach outperforms the existing methods.
\end{abstract}


\IEEEpeerreviewmaketitle

\section{Introduction}
Internet of Things (IoT) expects to connect massive devices for data collection and processing, which is considered as the enabling platform for a variety of promising applications, such as smart transportation and smart city~\cite{iots}. These IoT applications pose a high demand on storage and computing capacity, while the IoT devices are usually resource constrained. To process the data collected by the IoT devices, the data are typically sent to the remote cloud servers, where data processing and analysis are conducted for decision making. However, the cloud servers usually reside in the core network, which is far away from the data source. Moving massive data in and out of the remote cloud servers can cause potential traffic congestion and prolong the service latency. 

To address the above issues, mobile edge computing (MEC) emerges, which distributes computation and storage resources in proximity of the IoT devices~\cite{MEC1}\cite{asheralieva2020bayesian}. As a result, data generated by IoT devices can be offloaded to nearby MEC servers for processing, rather than sending data to the remote cloud servers~\cite{7120046}. By doing so, the potential congestion can be mitigated, and the service latency can be significantly reduced. Along with the benefits, MEC also encounters many challenges. Firstly, compared with cloud computing, MEC is usually with less capacity. Therefore, multiple MEC servers coordination is needed so that the tasks from IoT devices can be served efficiently. Secondly, some IoT services have stringent latency requirements, while service latency depends on many factors in offloading, such as transmission and computation power allocation. Thirdly, tasks with various service requirements are generated dynamically by IoT devices, and the MEC operating environment (e.g., channel conditions) changes over time, which requires the policy of the MEC system to be adjusted accordingly. Last but not least, energy consumption should also be considered, as IoT devices are energy constrained. 

In the literature, many approaches based on optimization techniques are proposed to allocate the MEC resources~\cite{MEC0,op0_survey,op1}, where different optimization problems are formulated and solved. However, the MEC system is usually very complex, and sometimes it is hard to be described in a mathematics form. Also, the optimization problem is mainly formulated based on a snapshot of the network, and it has to be reformulated when the condition changes over time. In addition, the majority of the classical optimization methods require a large number of iterations, and they may find a local optimum rather than the global optimum.  {\color{black} 
Deep Learning (DL)~\cite{DNN, DNN2} models that are remarkably successful in many supervised learning challenges such as computer vision, natural language processing, and self-driving cars, and DL models has be adopted to predict resource demands and optimize resource allocation to tackle the aforementioned challenges. Ale \etal~\cite{RNN} proposed a deep recurrent network to predict and update caching. However, standard supervised machine learning or DL methods require vast labeled datasets to train the models. It is considerably challenging to generate and label the data from the MEC network.

 In contrast, Reinforcement Learning (RL)~\cite{RL} does not require labeled data for training and the agent can learn the optimal policy through interacting with the MEC network environment. Therefore, RL has been employed in MEC to facilitate model-free control without knowing the internal transition of the system.} RL is utilized for MEC microservices coordination in~\cite{Q_learning}, and Q-learning is adopted for computing offloading control in~\cite{op2_Q}, and for optimizing remission rate and content delivery in~\cite{q_Spectrum}. Those standard RL methods leverage current and historical data for long-term decision-making, where rewards in both the current and the future time slots are considered. Classical reinforcement learning stores the learning results into a Q-table with tuples, including states, action, and values. However, they cannot be applicable when the state or action space is huge.




Deep Reinforcement Learning (DRL)~\cite{DRL,drl_humanlevel} can help address the above issues in MEC, by integrating neural network into reinforcement learning to approximate the Q values. In~\cite{drl_vecular}, a DRL based method is introduced to control computation offloading and minimize energy consumption for the Internet of Vehicles (IoV). To improve the Quality-of-Service (QoS), a resource allocation approach is proposed in~\cite{Li8906135} to minimize the service delay. Similarly, a Q-Network learning method is adopted in~\cite{Wang8303773} to maximize the number of successful transmissions. A DRL based collaborative MEC scheme is presented in~\cite{DRL_MEC1} to minimize the response latency and energy consumption. However, the action space is relatively small, where the agent can only take two actions, i.e., offloading to MEC servers or executing the task locally. An online DRL based offloading control approach for MEC is proposed in~\cite{DRL_MEC2}. Although both~\cite{DRL_MEC1, DRL_MEC2} aim to optimize multiple objectives, DRL is only for part of the problem and traditional optimization methods (e.g., linear programming) are used to deal with the rest based on the outputs of the DRL. Although some issues can be addressed using classical optimization methods, given the output from DRL, the optimization still only focuses on the current time step parameters. Moreover, the DRL models are only trained to optimize the partial target, which cannot fully exploit its advantages to learn the overall best policy or even squeeze other targets. 
{\color{black} Chen \etal~\cite{9085261} proposed a multiple-agent learning with Long Short-Term Memory (LSTM) to allocate resources for video stream in wireless network. Similarly, an asynchronous advantage actor-critic (A3C) based model was proposed to render offloading Virtual Reality (VR) video streaming~\cite{9120235}.}



In this work, we propose a deep reinforcement learning approach for delay-aware and energy-efficient offloading in a dynamic MEC network with multiple users and multiple MEC servers. The proposed learning scheme jointly optimizes the offloading servers and computational frequency allocation of edge servers to maximize the long-time utility, which incorporates both the number of completed tasks before their deadlines and energy efficiency. Unlike previous work, no optimization functions are required after the DRL model takes the actions. The offloading problem is tacked in an end-to-end manner using DRL, which can make relatively complex decisions based on the current information of the tasks and the network environment. All the optimization targets will be accomplished in one step by the DRL. In short, the main contributions of this work can be summarized as follow:
\begin{itemize}
  \item {\color{black}First, we propose an end-to-end Deep Reinforcement Learning (DRL) model to maximize the number of computational tasks before their respective deadlines and minimize energy consumption simultaneously. The proposed model can handle a relatively large action space and does not rely on standard optimization methods. Further, it can reduce the computational cost and make the optimal decisions at different states to maximize the expected long-term compensation.}
  
  
  \item {\color{black} Second, we capture the complexity of the MEC system by including time-varying channel conditions, various task profiles, and servers' state information into the state of the system; and propose model-free solutions for MEC systems. The agent cannot fully observe the environment states and is not aware of the internal transition mechanism. Besides, the observed states contain continuous variables such as channel gain. Moreover, the clip reward tricks that is regarded as a simple version of Clipped Surrogate Function~\cite{SchulmanWDRK17}  can prevent the model from suffering from oscillation during the training.}

  \item Finally, extensive simulations are conducted to evaluate the performance of the proposed approach. Simulation results demonstrate that the proposed approach can process more tasks before their deadlines while consuming less energy, compared than existing methods.
  
  
\end{itemize}

The rest of the paper is organized as follows. Section \Romannum{2} presents the system modeling and problem formulation. Section \Romannum{3} proposes the DRL model and training processes. Section \Romannum{4} provides the simulation and results analysis while Section \Romannum{5} concludes this work.

\section{System Model and Problem formulation}

\begin{table}[t!] \centering
\caption{Key Notations}
\begin{tabular}{p{3cm}|p{8cm} } 
 \hline
   \hline
$u_n^{(t)}$ & $n^{th}$ User at time slot $t$ \\ 
  \hline
$m_k^{(t)}$ & $k^{th}$ MEC server at time slot $t$ \\ 
  \hline
$T_{i,j}^{(t)}$ & $j^{th}$ task from $i^{th}$ user at the $t$ time slot \\
  \hline

$D_{i,j}$ & data size of task $\Omega_{i,j}^{(t)}$ \\
  \hline
  $C_{i,j}$ & demanded CPU cycles to process $T_{i,j}^{(t)}$ \\
  \hline
  
  $\Delta_{max}$ &  maximum tolerate time of task $\Omega_{i,j}^{(t)}$\\
  \hline
  $Q_k$  &  tasks queue in MEC server $m_k$\\
  \hline
   $f_{max}^{(k)}$ &  maximum frequency of MEC server $m_k$\\
  \hline
  $\delta_k^R$  &   remain running time of the current task running \\
  \hline 
  $\delta_{i,j}^T$  &  transfer time of task $\Omega_{i,j}$\\
  \hline 
    $\delta_{i,j}^C$  &  computing time of task $\Omega_{i,j}$\\
  \hline 
  $\delta_k^Q$  &  waiting for tasks in the queue time\\
  \hline 
  $\zeta_{i,j}$ &  transfer speed of the connection\\
  \hline 
  \hline
\end{tabular}
\end{table}

\subsection{System Model}
\begin{figure}[h!]
    \centering
    \includegraphics[width=6.0in]{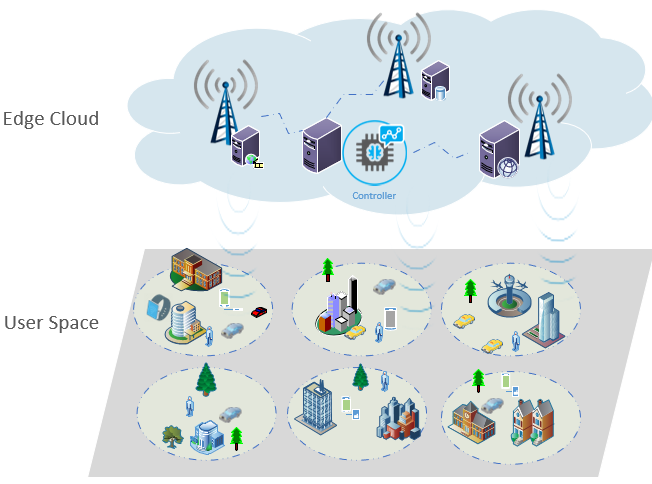}
    \caption{System Model}
    \label{fig:system_model}
\end{figure}

As shown in Fig.~\ref{fig:system_model}, there is a set of users $\mathcal{U}_t=\{u_1,u_2,\dots,u_N \}$ at a given time slot $t$, where $N$ is the number of users.
{\color{black} There also exists a set of MEC server $\mathcal{M}_t=\{ m_1,m_2,\dots,m_K \}$}, where $K$ is the number of MEC servers. MEC servers are with different computing capacities and the heterogeneity of MEC servers can be captured by the set of $\mathcal{M}_t = \{(Q_1,f_{max}^{(1)}),\dots,(Q_k,f_{max}^{(k)}),\dots,(Q_K,f_{max}^{(K)}) \}$, {\color{black} where $Q_k$ and $f_{max}^{(k)}$ denote the task queue and maximum CPU frequency of the $k^{th}$ server. The users can generate and choose to offload more than one task at a given time slot or process the tasks locally.} The set of tasks is denoted as $\Omega_t =\{\Omega_{0,0},\Omega_{0,1},\dots, \Omega_{i,j},\dots,\}$, where $\Omega_{i,j}$ represents the $j^{th}$ task of user $i$ and is defined as $\Omega_{i,j}=\{D_{i,j},C_{i,j},\Delta_{max}\}$; where $D_{i,j},C_{i,j},\Delta_{max}$ are the data size, the requested CPU cycles and the maximum tolerable time of the task. {\color{black}Suppose that there exists a coordinator/control agent which can communicate with the MEC servers. This coordinator can run on any of the local MEC or cloud servers to coordinate the MEC system. The control agent will collect the state information about the system, such as the tasks profile and status of each MEC server, and run a learning model to output the optimal decisions to execute the tasks. The action includes the index of the MEC server for offloading a certain task, and the CPU frequency at MEC servers to execute the tasks. Then, following the instructions from the control agent, the tasks will be offloaded to the selected MEC servers, and the servers process the tasks with the recommended CPU frequencies.}

The objective of the system is to maximize the number of tasks that are completed timely while minimizing the energy consumption. A task $\Omega_{i,j}$ is considered to be successfully processed if the total service time is less than its maximum tolerable time $\Delta_{max}$; otherwise the task fails, formally,
\begin{equation}
\label{eqn_time}
F_{task}=
\left\{
\begin{array}{ll}
    {1}, & {\text { if }\delta_k^R + \delta_k^Q + \delta_{i,j}^{T}+\delta_{i,j}^{c} \leq \Delta_{\text {max }}}; \\ 
    {0}, & {\text{ otherwise}}.
\end{array}
\right.
\end{equation}

The total service time for task $\delta_{i,j}$ includes the residual time $\delta_k^R$ of the current task running in the selected server, the transmission time to offload the task to the edge server $\delta_{i,j}^{T}$, the waiting time in the queue before service $\delta_k^Q$, and the computing time $\delta_{i,j}^C$. 
The residual running time $\delta_k^R$ of the current task being executed in the target server can be computed as the total computing time $\delta_{current}^C$ of the current running task minus the start running time $\delta_{run}$ for the current task, namely,
\begin{equation}
\label{eqn_waiting_run_time}
\delta_k^R = \delta_{current}^C - \delta_{run}.
\end{equation}
The waiting time for the task in the queue $\delta_k^Q$ is simply the summation of the computing time for all the {\color{black} tasks  before the current task in the queue}. The computing time for a particular task can be obtained by dividing the required CPU cycles $C_{*,j}$~\cite{cycles} by the recommended frequency $f_{*,j}$. Thus, the waiting time can be given by
\begin{equation}
\label{eqn_queue_time}
\delta_k^Q = \sum_{j=0}^M \frac{C_{*,j}}{f_{
*,j}}.
\end{equation}
The transmission time $\delta_{i,j}^T$ can be given as below
\begin{equation}
\label{eqn_tranfer_time}
\delta_{i,j}^T = \frac{D_{i,j}}{\zeta_{i,j}},
\end{equation}
where $D_{i,j}$ is the data size of the task and $\zeta_{i,j}$ is the transmission rate. The transmission rate $\zeta_{i,j}$~\cite{channel} can be 
given by
{\color{black}
\begin{equation}
\label{eqn_speed}
\zeta_{i,k} = \alpha_{i,l}^{(B)}log_2(1+\frac{P_{i,k}h_{i,k}L_{i,k}}{N_0}),
\end{equation}
}
where $\alpha_{i,l}^{(B)}$ is the bandwidth, $P_{i,l}$ is the transmission power, and $h_{i,l}, L_{i,l}$ are Rayleigh fading and path loss, respectively. Suppose that the task is offloading to the $k^{th}$, and the computing time $\delta_{i,j}^C$ can be calculated by 
\begin{equation}
\label{eqn_compute_time}
\delta_{i,j}^C = \frac{C_{i,j}}{f_{i,j}^k},
\end{equation}
where $C_{i,j}$ is the requested CPU cycles {\color{black} to compute task $\Omega_{i,j}$} and the recommended frequency $f_{i,j}^k$.

Moreover, the total energy consumption $E_{i,j}$ is sum of the energy consumption for transmission denoted by $E_{i,j}^T$ and computation energy consumption denoted by $E_{i,j}^C$, that is,
\begin{equation}
\label{eqn_energy}
E_{i,j} =E_{i,j}^T + E_{i,j}^C.
\end{equation}
Note that the energy consumption due to transmission is given by
\begin{equation}
\label{eqn_transfer_energy}
E_{i,j}^T = \delta_{i,j}^T P_{i,j} = \frac{D_{i,j}}{\zeta_{i,j}}P_{i,j}.
\end{equation}
The computation energy consumption~\cite{8016573} can be calculated by
\begin{equation}
\label{eqn_transfer_energy}
E_{i,j}^C =c(f_{i,j}^k)^2C_{i,j},
\end{equation}
where $c=10^{-26}$~\cite{7542156,8352664} and $f_{i,j}^k$ is the frequency used to compute the task $\Omega_{i,j}$ in the $k^{th}$ edge server.

\subsection{Problem Formulation}
This work focuses on the long-term utility of the system, and the objective is to maximize the number of tasks completed before the deadlines and minimize the energy consumption in the long run. To this end, we formulate this optimization problem as a Markov Decision Process (MDP) to maximize expected long-term rewards. 
Concretely, we consider an episode is terminated when one or more MEC servers are overloaded. Each episode contains many time slots denoted by $t$, and the reward $R_t$ depends on the action $a_t$ taken by the agent when the MEC network environment current state is $s_t$. The action space and state space are denoted by $A_t$ and $S_t$, respectively.




Ideally, we can distribute the tasks received at a time slot with a single action. However, the RL model can get diverged with explosion of action space when distributing multiple tasks from multiple users to multiple MEC servers with other action control parameters. The search action space would be growing exponentially as the number of users, tasks, or MEC servers increases. It is very challenging to guarantee that the agent can learn and converge to an optimal solution with the explosion of action space. 
{\color{black}
Let $N_\Omega$ be the number of received tasks, and $K$ be the number of MEC servers. Although it seems the action spaces equal to $K^{N_\Omega}$, the action space is far larger than the exponential function of the number of tasks because we have to consider the order of the tasks in each server. 



We can consider distributing identical tasks into distinct servers with different orders. We then can use permutation and combination function to derive the action space size as:
 \begin{equation}
\label{eqn_ac_size}
\begin{aligned}
A_s = P_{N_\Omega}^{N_\Omega} \times C^{(N_\Omega+1)+(K-2)}_{K-1} = N_\Omega!\times C^{N_\Omega+K-1}_{K-1}
\end{aligned}
\end{equation}
In our case, the model also recommends frequencies to run the tasks. Let $f_k$ be the action size of recommended frequencies; then the final action space can be given by

\begin{equation}
\label{eqn_huge_action}
A_s = N_\Omega !\times C^{N_\Omega+K-1}_{K-1}  \times f_k.
\end{equation}

}
Moreover, if tasks are processed in parallel, we need an immense input size for the learning model to receive a large number of tasks in peak-hours. However, in off-peak hours, the input feature for the learning model would be considerably sparse. Consequently, the learning model can probably process matrices full of zeros and waste the MEC resources most of the time.

To address the aforementioned challenges, we process the received tasks sequentially within a time slot.  Concretely, we assume that the channel distribution and other parameters of the MEC remain the same during time slot $t$, and the transition of states only depends on the actions taken by the control agent. Further,  the controller of the MEC network can collect profiles of the receive tasks in a queue, and the proposed learning agent can take actions with respect to task queue within $\tau$ time step, where $\tau << t$. Similarly, the transition probability can be formed based on $\tau$ time step rather than time slot $t$ (Eq.\ref{eqn_state_transition_prob2}). The state transition with time step $\tau$ satisfies the property of MDP, and the transition probability  $p(s^{\prime} | s, a)$ is given as follows:

\begin{equation}
\label{eqn_state_transition_prob2}
p(s^{\prime} | s, a) \doteq \operatorname{Pr}\{s_\tau=s^\prime | s_{\tau-1}=s, a_{\tau-1}=a \}.
\end{equation}

In the following, we will present the details of the MDP formulation in the MEC network environment with the time step $\tau$ within the given time slot $t$. The states at time slot $t$ is given by 
\begin{equation}
\label{eqn_mec_state}
s_t = \{s_0, s_1, s_{\tau},\dots, s_{\lambda}|\sum_{\tau=0}^\lambda \tau << t \},
\end{equation}
where the $\lambda$ is the number of tasks received at $t$. Note that $s_\tau=\{\Omega_\tau,M_\tau,\zeta_\tau\}$, where $m_k^{(\tau)}$ is the $k^{th}$ MEC server state in $M_\tau$, and $\zeta_\tau$ is the transfer speed matrix to reach $\zeta_\tau$ servers.

We assume the number of tasks received by the MEC servers at time slot $t$ is less or equal than a threshold $\lambda$. {\color{black}The tasks with indices larger than $\lambda$ will be processed in the next time slot.} 
Moreover, the waiting time of the target server is updated when the model takes action because the tasks will be added to the queues at the edge servers. 

As shown in Section \Romannum{2}, each element of a general task queue in a MEC server is a pair consistent with the task information and recommended frequency. Similarly, we can derive the task queues in the $k^{th}$ MEC server with tasks and the recommended frequency $f_r^{(k,\tau)}$ to run the task 
\begin{equation}
\label{eqn_mec_server_queue}
Q_k ^\tau= \{(\Omega_{1,1}^{(1,\tau)}, f_r^{(1,\tau)}),\dots,(\Omega_{i,j}^{(k,\tau)}, f_r^{(k,\tau)}),\dots\},
\end{equation}
where $\Omega_{i,j}$ is the task from $j^{th}$ task from the $i^{th}$ user and start to distributed at time step $\tau$.


Similarly, we can denote the action during time slot $t$ with a set of actions that the agent can take within $t$, as

\begin{equation}
\label{eqn_mec_action_tau}
a_t = \{a_0, a_1, a_{\tau},\dots, a_{\lambda}|\sum_{\tau=0}^\lambda \tau << t \}.
\end{equation}
The actions that the agent can choose include the index of edge servers for offloading tasks and the recommended CPU frequency for edge servers. Therefore, the size of the action space is $K\times f_p$, where $K$ is the number of MEC servers, and $f_p$ is the resolution of discrete percents of the maximum CPU frequency of the edge server. A general action can be defined as $a_\tau = (k,f_r)$, where $k$ is the target offloading server, and $f_r$ is a recommended percentage of maximum frequency to execute the given task. The possible values of  $f_r$ can range from 0\% to 100\% .

As the final component of the MDP framework, $R_t(s_t,a_t)$ is the reward obtained in time slot $t$. Similar to the state and action, the reward at time slot $t$ is a collection of rewards $R_\tau(s_\tau, a_\tau)$ within $t$, i.e., 
\begin{equation}
\label{eqn_reward_time_slot}
R_t(s_t,a_t) = \sum_{\tau=0}^\lambda  R_\tau(s_\tau, a_\tau),
\end{equation}
with $\sum_{\tau=0}^\lambda \tau << t$.

{\color{black}
Theoretically, we can formulate the following optimization problem as:
\begin{equation}
\label{reward0}
\begin{aligned}
\min_{a_\tau} \quad & \sum_{\tau=0}^N E_\tau^T + E_\tau^C\\
& = \sum_{\tau=0}^N \left[ \frac{D_{i,j}}{\zeta_{i,j}}P_{i,j} + c(f_{i,j}^k)^2C_{i,j}\right] \\
\textrm{s.t.} \quad & \delta_k^R + \delta_k^Q + \delta_{i,j}^{T}+\delta_{i,j}^{c} \leq \Delta_{\text {max }},\\
& f_{i,j}^k \leq f_k^{max},\\
& a_\tau = \{k,f_{i,j}^k\}
\end{aligned}
\end{equation}
where $E_\tau^T$ and $E_\tau^C$ are the energy cost for transmission and computation, respectively; $f_k^{max}$ is maximum frequency of the $k^{th}$ MEC server. However, there are several drawbacks in above-mentioned reward formulation. First, the constraint $\delta_k^R + \delta_k^Q + \delta_{\tau}^{T}+\delta_{\tau}^{c} \leq \Delta_{\text {max }}$ may not always satisfied, and the control agent has to deal with the cases when the no solution existing in the feasible areas. Second, it is not flexible to balance the energy cost and response delay. Third, the computational cost grows exponentially with the increase of the variables and the scale of the problem. Therefore, we design a flexible reward function that allows the DRL model to solve all the optimization target at one time in an end-to-end manner, as mentioned earlier.   
} Concretely, $R_\tau(s_\tau, a_\tau)$ is the reward at the time step $\tau$ indicating the number of the completed tasks and energy cost,
\begin{equation}
\label{eqn_reward_imidiate}
 R_\tau(s_\tau, a_\tau) = (1-\eta)\beta_1 F_{task} - \eta \beta_2 \log_2(E_\tau) + \mathcal{C},
\end{equation}
where $\beta_1$ and $\beta_2$ are the terms for normalizing the profit from completing tasks on time and the energy cost to the same scale.  $\eta \in [0,1]$ is a weight to balance the number of tasks completed and the energy consumption, which can be adjusted based on different applications. $\mathcal{C}$ term is a positive constant number that to increase the accumulative {\color{black} rewards} with the number of time slots. The total accumulative reward contributed by $\mathcal{C}$ is equal to the number of time slots times $\mathcal{C}$.  In other words, the agent would find best policy to stop the severs from overloaded so that it can prolong the episodes and maintain the MEC network stability. Additionally, the energy cost term is scaled with logarithm because it is approximately proportional to the square of the MEC running frequencies and considerably fluctuates.

Although the goal is to maximize expected long-term rewards, the learning agent can only have immediate reward from the current time step, and rewards of future time steps are unknown. To evaluate the current action on the long-term reward, we utilize both the immediate reward and the expected rewards of the future estimated with learned policies. 
Specifically, the current action is evaluated by long-term return:
\begin{equation}
\label{eqn_return}
G_\tau \doteq R_{\tau}+\gamma R_{\tau+1}+\gamma^{2}R_{\tau+2} +\cdots=\sum_{k=0}^{\infty} \gamma^{k} R_{\tau+k},
\end{equation}
which contains the immediate reward and the discounted further rewards. $0<\gamma<1$ is the discount factor. The immediate reward can be feedback from the environment, and the expected future rewards are computed with a policy $\pi$ from the trained model. A policy $\pi$ is a set of actions that the agent follows to interact with the environment.

Therefore, the goal of this work is to develop a learning model to find optimal policies $\pi^*$ such that  action-value function is maximized:
\begin{equation}
\label{eqn_policy}
Q^{*}(s, a)=\max _{\pi} \mathbb{E}\left[R_\tau+\gamma G_{\tau+1} | s_\tau=s, a_\tau=a, \pi\right],
\end{equation}
where $R_\tau$ is the immediate reward, and $G_{\tau+1}$ is the expected future reward with discounted by $\gamma$.

\section{Proposed Method}

In this section, we present the proposed DRL method to maximize the number of completed tasks and minimize the system energy consumption by dynamically {\color{black} determining the MEC servers for offloading and the computational frequency allocation. Specifically, the proposed DRL model can learn and generate optimal policies that maximize} the long-term reward. By inputting observed data from the MEC network, the DRL model produces control parameters to maximize the number of completed tasks and minimize energy consumption. {\color{black} In what follows, we will present the data preprocessing,  the DRL model, and the training process.}

\subsection{Data Prepossessing}
Data preprocessing is a critical component of the proposed method. The data is considerably complicated and noisy, which increases the training effort for the model. The raw {\color{black} features}, including the channels, server states, and users' tasks, are time-varying. Moreover, since the scale ranges of raw features are significantly different, the proposed models might ignore the essential features. Finally, the dimension of input features is considerably high because the features contain the channel distributions into the states, and the number of channels is increasing with the number of users and MEC servers. Without the data preprocessing, the DRL agent possibly overlooks the essential features if it is fed raw data, which can cause the agent {\color{black} to converge too  slowly for optimal solutions or converge to non-optimal solutions.

To deal with the above issues, we adopt normalization methods to rescale and concatenate the features in a desirable format. The features consist of hierarchical components, and are stored in a tree-like data structure.} For instance, we have a root node, and a branch represents features from the MEC servers; further, the branch has three sub-branches including MEC servers' state $M_\tau$, transfer speed matrix $\zeta_\tau$, and the queue tasks $\Omega_\tau$, and each of them contains some leaf-level nodes. Therefore, we have to normalize the leaf-level sub-components and concatenate them together. We first compute the Frobenius norm (Eq.\ref{eqn_normalize}) for all the leaf-level components of the feature $\mathbb{R}^{m \times n}$~\cite{normalize}. We then compute the normalization of the matrix $\mathbb{R}^{m \times n}_N$ by dividing the normal in an element-wise manner. That is, 
\begin{equation}
\label{eqn_normalize}
\begin{aligned}
\|A\|_F=\sqrt{\sum_{i=1}^m \sum_{j=1}^n\left|a_{i j}\right|^{2}},
\end{aligned}
\end{equation}
where $a_{ij}\in \mathbb{R}^{m \times n}$ with
\begin{equation}
\label{eqn_normalize1}
\mathbb{R}^{m \times n}_N= \frac{\mathbb{R}^{m \times n} }{\|A\|_F}.
\end{equation}
Finally, we concatenate all the normalized sub-features as a single feature, which is ready to feed to the learning model.

\subsection{DRL Model}

In this subsection, we propose a DRL model to address the joint optimization problem formulated in the previous section. The MEC network is regarded as the Reinforcement Learning (RL) environment, and the proposed DRL mode as a learning agent that can interact with the MEC network and learn from the experience. Due to the complexity of the MEC network environment, it is almost impossible that the states and transitions are fully observable to the agent. For the simplicity of argument, we consider the MEC network environment as a MDP with internal transition probability $P(r,s\prime|s, a )$, which is opaque to the DRL agent. Further, we propose a model-free DRL model to learn from the environment without knowing its internal transition. The DRL agent is expected to generate robust policies that can maximize long-term accumulated rewards.

\begin{figure*}[h!]
    \centering
    \captionsetup{justification=centering}
    \includegraphics[width=6in]{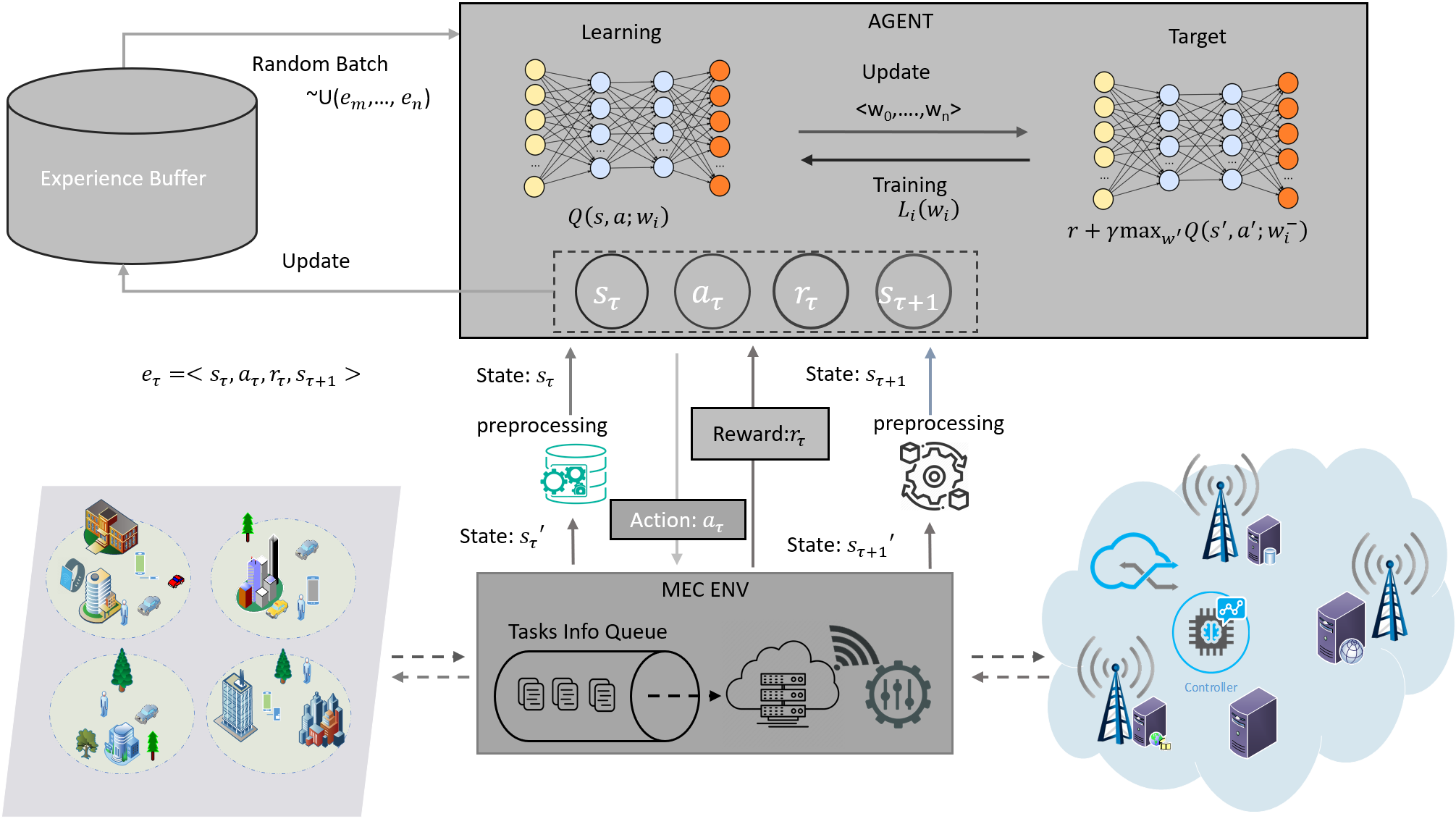}
    \caption{Offloading System}
    \label{fig:offlaoding_system}
\end{figure*}
\subsubsection{Reinforcement learning Framework} 
Reinforcement Learning (RL) is a method that allows a learning agent to learn by interacting and exploring the unknown environment. Unlike standard machine learning, the RL models can learn from the sequential and evaluative feedback from the environment.

To learn from the unknown MEC environment, the RL agent is required to balance exploitation and exploration. Exploitation is to capitalize the learned knowledge by greedily exploring search space with respect to Q-value, namely,
\begin{equation}
     \label{eqn_off_policy}
    \begin{aligned} 
    a=\argmax_{a^\prime} Q\left(s, a^\prime ; w \right),
    \end{aligned}
\end{equation}
where $w$ is the parameters matrix. {\color{black}On the other hand, exploration allows the learning agent to acquire knowledge about the MEC network environment by taking actions randomly.}
In this study, we adopt the $\epsilon-greedy$ method to balance exploitation and exploration, which means the model selects actions with a greedy algorithm with probability $1-\epsilon$ and randomly selects actions with probability $\epsilon$. Initially, the agent has no knowledge of the MEC network environment, and it takes more random actions in the early episodes to explore the environment. As the agent gradually acquires enough knowledge about the environment, the agent starts to exploit the learned knowledge to generate optimal policies. Therefore,  $\epsilon$ is designed to decrease over the episodes.

Moreover, the goal of the RL agent is to derive the optimal policy $\pi^*$ by finding the optimal action-values $Q^*(s,a)$ that maximizes the long-term accumulative rewards. Action-value $Q(s,a)$  is generated by  taking action $a$ state $s$, and then follows with the policy $\pi$. The optimal action-value $Q^*(s,a)$ is the maximum value of all possible values of $Q(s,a)$, i.e., 
\begin{equation}
\label{eqn_max_Q_pi}
\begin{aligned} 
Q^{*}(s, a)=\max_{\pi} \mathbb{E}\left[R_\tau | s_\tau =s, a_\tau=a, \pi\right].
\end{aligned}
\end{equation}
The policy that can optimize the action-value $Q^*(s,a)$ is the optimal policy $\pi^*$ and satisfies the
%
Bellman equation
\begin{equation}
\label{eqn_bell_man}
\begin{aligned} 
Q(s, a)=\mathbb{E}_{s^{\prime}}\left[R+\gamma \max _{a^{\prime}} Q\left(s^{\prime}, a^{\prime}\right) | s, a\right].
\end{aligned}
\end{equation}
The main idea of deriving optimal action-value (aka, Q-value) is to take action $a^\prime$ from all possible actions for the next step of $Q(s, a)$ that maximizes $R+Q(s^\prime, a^\prime)$, and repeat this step over all the states to generate optimal policies. Theoretically, we can derive the optimal action-value by updating the Bellman equation iteratively.  $Q(s, a)$ value keeps improving over the iteration, and the  $Q_\tau \rightarrow Q^*$ as $\tau \rightarrow \infty$, as given by,
\begin{equation}
\label{eqn_bell_inter}
\begin{aligned} 
Q_{\tau+1}(s, a)=\mathbb{E}_{ s^\prime}\left[R+\gamma \max _{a^{\prime}} Q_{\tau}\left(s^{\prime}, a^{\prime}\right) | s, a\right],
\end{aligned}
\end{equation}
where $Q_\tau$ is the Q-value at step $\tau$, and $Q^*$ is the optimal value-function.

However, finding an optimal Q-value by iterating over $\infty$ times is impractical in real-world applications. Consequently, the classical RL models would diverge from finding an optimal policy in a vast or continuous search space because it is nearly impossible for the agent to explore all of the search space. Fortunately, we can reduce the search space by approximate functions, including linear and non-linear functions. A deep neural network can be considered as a non-linear function that can approximate many complex states, $Q(s, a; w) \approx Q^*(s, a)$.

\subsubsection{Deep Reinforcement Learning Model}

With the complexity and continuous states, it is almost impossible to store all the state-action value pairs in a Q-table that allows standard RL methods to search the optimal policies. Although we could turn the continuous space into discrete space through discretization of the continuous space, it is challenging to balance the resolution of discrete space. {\color{black} On the one hand, the low-resolution discretization of space compromises the accuracy of the representation of the features. On the other hand, a high-resolution discrete space would generate a vast search space that increases search time and complexity; further, enormous search space hinders the model from convergence and finding the optimal policies.} Therefore, we adopt a deep neural network as an approximator to represent the search space. Specifically, the deep neural network represents input states, and the model computes probabilities of all possible actions $P_A = \{p_1,\dots,p_k\}$ at one time. The agent selects the actions based on the probabilities and interactions with the environment.

The DRL agent is the backbone of the proposed method. To illustrate the proposed DRL model, we present the offloading system, as shown in Fig.~\ref{fig:offlaoding_system}. First, a coordinator is placed in the MEC network; the coordinator collects the state information of the system and provides an interface to the DRL agent. Specifically, the coordinator first collects offloading tasks' profiles and places them into the queue in the MEC environment (MEC ENV). Second, the DRL agent takes action based on the state information observed from the MEC environment. Third, the coordinator in the MEC environment executes the action by offloading tasks to the target server, and then the MEC servers run the tasks with frequencies recommended by the DRL agent.  Fourth, the DRL agent stores the data (state, action, reward, and next state) into the experience replay buffer for training the DRL model. {\color{black} Fifth, the DRL agent draws sample data from the experience replay buffer and trains the learning network by minimizing the loss function defined by a Mean Square Error (MSE). Finally, the target network is updated after every $\mathcal{N}$ episodes.} The training steps can entirely separate from the above steps, which means we can run the training process with the above steps simultaneously.

In the proposed DRL model, the approximate function is a neural network (also known as Q-Network) with parameters $w$. {\color{black} Incorporating deep neural networks with RL is considerably difficult to train, as the unknown fluctuates feedback from the dynamic environment. In order to mitigate the oscillation and prevent divergence during training, the deep Q-network method introduces fixation methods. Specifically, the approximator neural network has a copy with fixed parameters $w^-$, also known as the target network, where the weights keep unchanged in a certain number of the episode. The other copy of the neural network parameters $w$ called the primary network (also known as local learning network) keeps learning from the data in the memory buffer, and its weights are copied to the target network after every $\mathcal{N}$ episodes.} The parameters $w_i$ are updated to minimize the loss function, which is the MSE between current action-value with $Q(s^\prime, a^\prime; w_\tau)$ and optimal $Q^*(s^\prime, a^\prime )$, which can be substituted with fixation term $\Bar{F}$ 
\begin{equation}
\label{eqn_fixation}
\begin{aligned} 
\Bar{F} = r+\gamma \max _{a^{\prime}} Q^{*}\left(s^{\prime}, a^{\prime};w^- _\tau \right),
\end{aligned}
\end{equation}
%
to derive the loss function 
\begin{equation}
\label{eqn_loss}
\begin{aligned}
 L_\tau(w_\tau) &=\mathbb{E}_{ s,a,r}\left[\left(\mathbb{E}_s[\Bar{F}| s, a]-Q\left(s, a ; w_\tau \right)\right)^2\right]. 
\end{aligned}
\end{equation}
where $w^-_\tau$ is updated in previous iterations. The complete algorithm can be found in Alg.~\ref{agl:model_train}.

\subsection{Training Process}
In this subsection, we present the training process of the DRL model. To simplify the illustration, we present the training process with three components: initialization and preparation, generating training data, and learning from the data. Also, the techniques introduced in the training process is provided in the following.

\begin{algorithm}
    \KwIn{$epoch\_no, \epsilon_{start},\epsilon_{end}$}
    \KwOut{$loss, gradients$}
    //1. Initialization: \\
    Initialize replay memory $D$ with capacity $N$; \\
    Initialize action-value $Q$ with random weights $w$; \\ 
    Initialize target action-value $\hat{Q}$ weights $w^- \leftarrow w$; \\
    Initialize scores with window size;\\
    $\epsilon \leftarrow  \epsilon_{start}$; \\
    \For{$episode \leftarrow 1\ to \  M$}
    {
        Initialize input raw data $x_1$;\\
        Prepossess initial state: $S \leftarrow \phi(<x_1>)$;\\
            \For{time step: $\tau \leftarrow 1\ to \ T_{max}$}
            {
                // 2. Generate training data:\\
                Select action A from state S using: $\pi \leftarrow \epsilon-Greedy(\hat{Q}(S,A,w))$; \\
                Take action A, Observe reward R and get next input $s_{\tau+1}$; \\
                Prepossessing next state: $S^\prime \leftarrow \phi(s_{\tau+1})$;\\
                Store experience tuple $(S,A,R,S^\prime)$ in replay memory $D$; \\
                $S^ \prime \leftarrow S$; \\
                // 3. Learning: \\ 
                Obtain random mini-batch of $(s_j,a_j,r_j,s_{j+1})$ from $D$;\\
                \eIf{episode terminate at step $j+1$}
                {
                Set target $\Bar{F}_j \leftarrow  r_j$; \\
                }{
                    Set target $\Bar{F}_j \leftarrow  r_j+\gamma \max_{a^{\prime}} \hat{Q}\left(s_j, a,w^- \right)$; \\
                }

                Update: $w \leftarrow  w + \alpha \nabla_{w_j}L\left(w_j\right)$ with $Adam$; \\
                Every $\mathcal{N}$ steps, update: $w^- \leftarrow  w$; \\
            }
            $\epsilon \leftarrow  max(\epsilon_{end},\epsilon*decay)$; \\
            Store score for current episode;\\
    }

\caption{DQN-Learning for MEC \label{agl:model_train}}
\end{algorithm}

\subsubsection{Initialization}

As shown in Alg.\ref{agl:model_train}, the algorithm {\color{black}starts with initializing the experience replay memory buffer, the exploration proportion $\epsilon$, and two neural networks (i.e., learning and the target network). The replay buffer stores} experience of the DRL agent when interacting with the MEC network environment. A learning network is initialized with random weights and replicated to the target network. Then, the algorithm iteratively generates data and trains the DRL model over the episodes. The episode ends when the time step is larger than or equal to the threshold $T_{max}$, or the environment returns a finish flag, which indicates at least one of the MEC servers is overloaded. 

\subsubsection{Exploration and Data Acquisition}

The DRL agent interacts with the MEC network environment to generate the training dataset. Concretely, the DRL agent acquires knowledge by using a $\epsilon-greedy$ method, as mentioned before. The agent randomly explores the environment and produces greedy actions with a probability of $\epsilon$, and then it selects other actions with a probability of $1-\epsilon$. Each interaction generates {\color{black}a tuple containing the} current state $s_\tau$, action $a_\tau$, reward $r_\tau$ and s$_{\tau+1}$. Further, the generated data is saved into the experience buffer for learning purposes. Additionally, the data generation module and the learning module do not depend on each other; {\color{black} therefore, these two modules do not need to follow each other step by step. For example, for the learning module, the model can have several runs of the data generation or a single run; and they can run separately and simultaneously.}

\subsubsection{Replay Experience Buffer}

The sequences of the {\color{black} experience-tuples could be highly correlated when the agent interacts with the MEC network environment if the data is fed into the model sequentially. The classical Q-learning methods learning from sequentially ordered data cause risks of being swayed due to correlation among data. To prevent action values from oscillating or diverging, we adopt experiences replay method to draw training data uniform randomly from the experiences buffer. With this approach, }instead of learning data timely as interacting with the environment, the agent collects experiences tuples $<s, a,r,s^\prime>$ into the experiences buffer. The experience buffer is a queue with a fixed size, and the generated data is continually added into the queue. The experience buffer would delete the oldest data to make room for new data when it is full. With experiences replay and random samples, the actual loss function can be given as
\begin{equation}
\label{eqn_loss_uniform}
\begin{aligned} 
L_\tau(w_\tau)=\mathbb{E}_{(s, a, r, s^\prime) \sim U(D)}
\Bigl[\Bigl(r+\gamma \max _{a^\prime} Q(s^\prime, a^\prime; w_\tau^-)
-Q(s, a ; w_\tau)\Bigr)^2 \Bigr].
\end{aligned}
\end{equation}

The experience buffer introduces several enhancements to the training process. First, {\color{black}the correlation of the sequential order can be decoupled by sampling the training data from the experience buffer instead of feeding the sample one by one. Second, the experience replay allows the agent to learn more from individual entry multiple times. More importantly, the experience replay can recall rare occurrences to prevent the model from overfitting due to bias of training sample distribution. Third, it can mitigate the oscillation or divergence caused by outlier training samples by using batch samples. The model is allowed to sample multiple data samples to leverage the batch normalization to reduce the swaying. }

\subsubsection{Learning}

The DRL learning process is slightly different from training a conventional deep learning model. In the forward propagation, the model draws a batch of training samples from the experience buffer and {\color{black}feeds them to both the learning and target networks}; and then the loss (Eq.\ref{eqn_loss}) is computed with the errors between the rewards from the learning and target networks. Further, the parameters of the local learning network are updated with backpropagation, which has no significant difference {\color{black} when compared to} training a regular neural network. Therefore, we will not reiterate the details of this step. However, unlike the loss function of the standard neural network computed as the error between the outputs and labels, the loss function of DRL is computed by the outputs from the learning network and target network. The loss functions of DRL is computed by the {\color{black} difference between the outputs from the learning network and the target network because the }DRL agent learns from evaluative feedback rather than true label data. In brief, the loss is computed in the forward propagation, and then the parameters $w$ are adjusted $w$ with learning rate $\alpha$ times partial derivative loss function with respect to $w$, as follows:

\begin{equation}
\label{eqn_update_theta}
\begin{aligned}
w \leftarrow Adam(w, \alpha \nabla_{w_\tau}L\left(w_\tau\right)).
\end{aligned}
\end{equation}

To simplify the partial derivative of loss function, we first can derive the loss function as
\begin{equation}
\label{eqn_loss_final}
\begin{aligned}
 L_\tau(w_\tau) &=\mathbb{E}_{s, a, r, s^\prime}\left[\left(\Bar{F}-Q\left(s, a; w_\tau \right)\right)^2\right] + \mathbb{E}_{s, a, r}\Bigl[\mathbb{V}_s[\Bar{F}]\Bigr]. 
\end{aligned}
\end{equation}
Since the last term $\mathbb{E}_{s, a, r}\Bigl[\mathbb{V}_s[\Bar{F}]\Bigr]$ of the loss function does not depend on learning network parameters $w$, we  can ignore it during computing partial derivative with respect to $w$. {\color{black} In other words, we can derive the partial derivative of loss function by: 
\begin{equation}
\label{eqn_loss_dirivative0}
\begin{aligned}
\nabla_{w_\tau}L\left(w_\tau\right) &=\nabla_{w_\tau}\mathbb{E}_{s, a, r, s^\prime}\left[\left(\Bar{F}-Q\left(s, a; w_\tau \right)\right)^2\right] + \nabla_{w_\tau}\mathbb{E}_{s, a, r}\Bigl[\mathbb{V}_s[\Bar{F}]\Bigr] \\
&= \nabla_{w_\tau}\mathbb{E}_{s, a, r, s^\prime}\left[\left(\Bar{F}-Q\left(s, a; w_\tau \right)\right)^2\right];
\end{aligned}
\end{equation}
further, we can derive gradient of $L\left(w_\tau\right)$ by using the chain rule and substituting $\Bar{F}$ with $\Bar{F} = r+\gamma \max _{a^{\prime}} Q^{*}\left(s^{\prime}, a^{\prime};w^- _\tau \right)$. Therefore, the gradient of $L\left(w_\tau\right)$ is:
}
\begin{equation}
\label{eqn_loss_dirivative}
\begin{aligned}
\nabla_{w_\tau}L\left(w_\tau\right)=\mathbb{E}_{s, a, r, s^{\prime}}\Bigl[\Bigl(r+\gamma \max _{a^{\prime}} Q\left(s,^{\prime}, a^{\prime} ; w_\tau^{-}\right)  -Q\left(s, a ; w_\tau\right)\Bigr) \nabla_{w_\tau} Q\left(s, a ; w_\tau\right)\Bigr].
\end{aligned}
\end{equation}
where $w_\tau^-=w_{_\tau-1}$.
Additionally, we adopt Adam~\cite{adam} optimization function when updating the parameters. Finally, the updated parameters of learning network $w$ are copied to the target network parameters every $\mathcal{N}$ episodes to overwrite the $w^-$.  

\subsubsection{Reward Clipping}
To facilitate convergence and generate the optimal policies smoothly, {\color{black} both the rewards and the loss errors can be clipped}. Due to the complexity and uncertainty of the MEC network, the rewards obtained are significantly different even with a small change of the feature. Also, some features, such as channel distributions, may distort the DRL model back and forth as the features have a wide range and high variance of the training samples. Consequently, the DRL model can be slow to learn to converge, or it may never converge to the {\color{black} optimal polices}. The clip is a straightforward but practical technique that can mitigate those issues. With clips, any element in  $\mathbb{R}^{m \times n}$ less than $min$ would be replaced with $min$, and greater than $max$ would be replaced with $max$

\begin{equation}
\label{eqn_clipp}
\begin{aligned}
 \mathbb{R}^{m \times n}_{clipped} = clip(\mathbb{R}^{m \times n},min,max).
\end{aligned}
\end{equation}

\section{Simulation Results}
In this section, we present the simulation results. We adopt Python as the programming language in this simulation and use processes to simulate the entities, including users, MEC servers, the control agent in the MEC network, and the DRL agent. Additionally, we choose PyTorch\footnote{https://pytorch.org} and NumPy\footnote{https://numpy.org} to build a DRL model to reduce implementation efforts. We run the simulation with various parameter settings to compare the results and verify the proposed model. 

According the architecture of the proposed method shown in Fig.\ref{fig:offlaoding_system}, the simulation system contains two major parts, {\color{black} the DRL agent and the MEC network environment (MEC ENV). The MEC ENV is made of three components: the users, MEC servers (Local Base stations), and coordinator. Additionally, the MEC network also maintains various states, such as the channel signal distribution and speed distribution.} Some of the critical simulation parameters are summarized in Table.~\ref{tabke:para}.  First, the users are the task generators, and each user can create multiple tasks by waiting a random time period (around 0.001 seconds). Second, the simulator generates a set of MEC servers based on the parameter settings, including minimum and maximum frequencies, sizes of task queues, and overload thresholds. The MEC servers also maintain status information, process the tasks, and compute the rewards. {\color{black} The reward is related to the number of tasks completed before their tolerant time and minimize energy consumption. Specifically, a MEC server runs with the recommended frequency when it receives a task, and then the reward can be computed based on the  recommended frequency and required CPU cycles. The MEC is in the idle status with minimum frequency when the task queue is empty. }

The DRL agent maintains the replay buffer and two neural networks, namely the local learning network and the target network. The DRL agent interacts with the MEC network through the coordinator, then learns and generates actions and policies for the MEC network. In this simulation, the neural networks have five hidden layers, where the number of the neurons from layer one to layer five are 256, 512, 512, 512 {\color{black}and 256. Other parameters are given in Table.\ref{tabke:para}.  Each neural network has an input layer and an output layer, and the numbers of neurons are equal to the sizes of state space and action space, respectively.}   

\begin{table}[t!] \centering
{\color{black}
\caption{Parameter Setting}
\label{tabke:para}
\begin{tabular}{p{3.5cm}|p{8.5cm} } 
 \hline
   \hline
Signal to Noise Ratio (dB) & 100 \\ 
  \hline

Task Data Size (bits)& $[2 \times 10^5,2 \times 10^7]$\\ 
  \hline
Task Computing Size (cycles) &  $[8 \times 10^6, 1\times10^7]$ \\ 
  \hline  
Server max frequency (Hz) & $[2\times10^9,8\times10^9]$ \\ 
  \hline  
Number of Online User & $[10,1000]$ \\ 
  \hline    
Batch Size & 256\\ 
  \hline  
Learning Rate $\alpha$ & $5\times 10^{-4}$\\
  \hline 
Discount Factor $\gamma$ & 0.9\\ 
  \hline   
  \hline
\end{tabular}
}
\end{table}

\begin{figure}[h!]
  \centering
  \begin{minipage}[b]{0.4\textwidth}
    \includegraphics[width=3.0in]{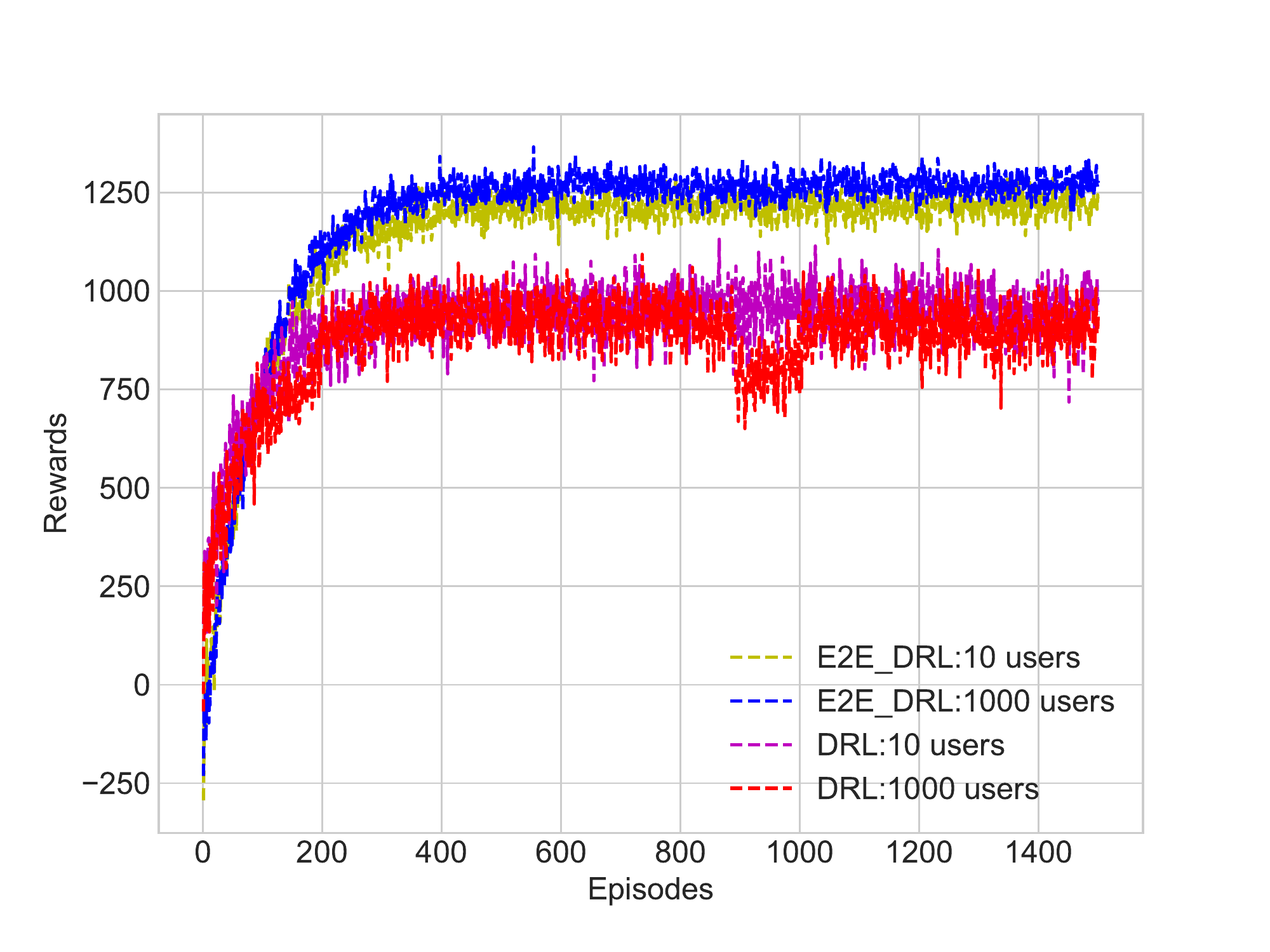}
    \caption{Learning curve}
    \label{fig:learning}
  \end{minipage}
  \hfill
  \begin{minipage}[b]{0.4\textwidth}
    \centering
    \includegraphics[width=3.0in]{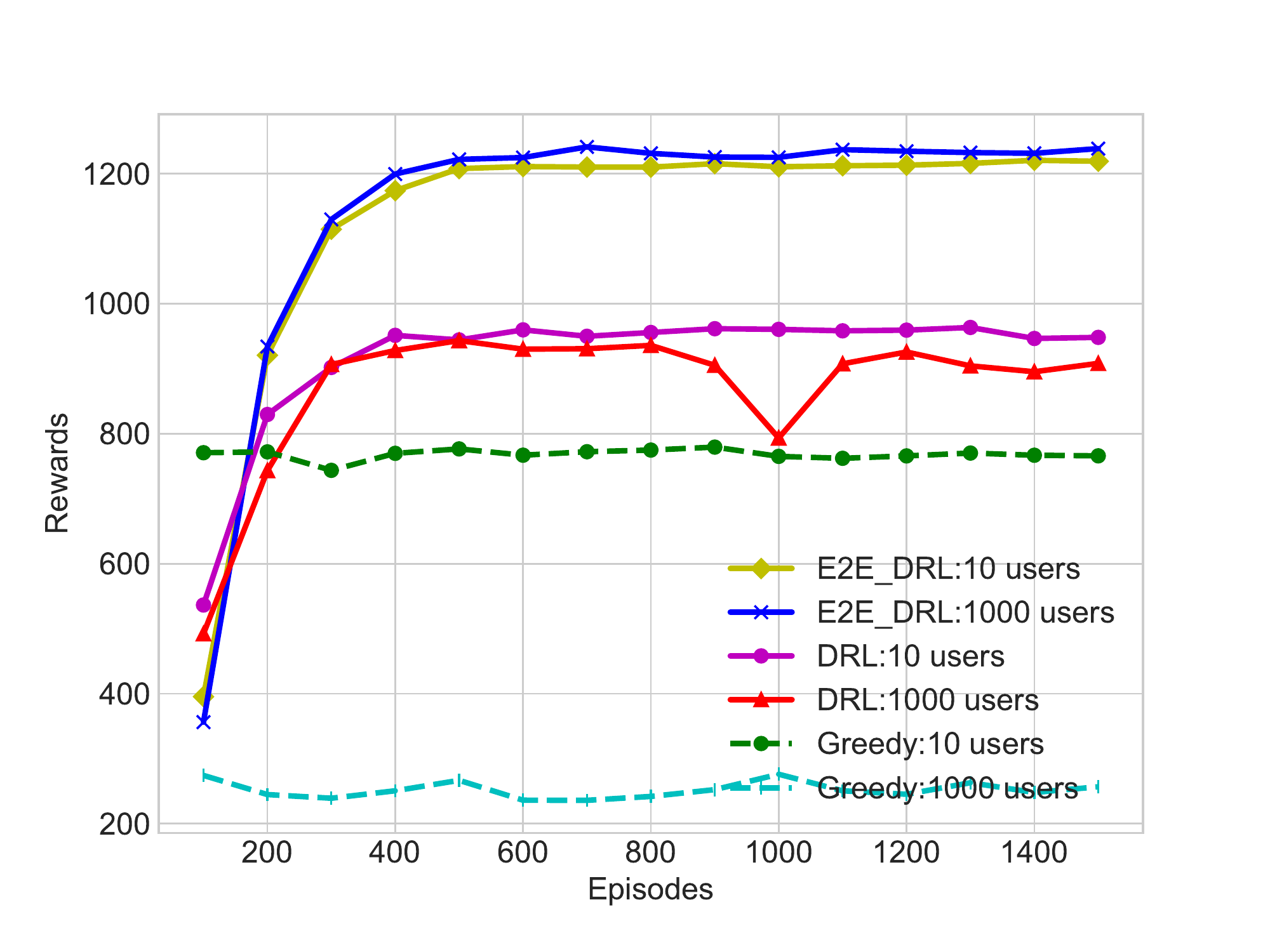}
    \caption{Reward comparison }
    \label{fig:reward_cmp}
  \end{minipage}
\end{figure}

{\color{black}Fig.~\ref{fig:learning} shows the performance of proposed End-to-End DRL (E2E\_DRL) model and existing DRL models; the legends \textit{E2E\_DRL:10 users} and \textit{E2E\_DRL:1000 users}  denote the proposed E2E\_DRL model with 10 and 1000 users; and similarly, \textit{DRL:10 users} and \textit{DRL:1000 users}, denote the existing DRL models with 10 and 1000 users. As we can see from the figure, the models converged around 500 episodes and the proposed model can achieve more rewards than existing DRL models. The learning curves are swaying along with the episode because the generated data has many random factors, including the uncertainty of tasks and the MEC network environment. The existing DRL models have larger variances and relatively lower rewards than the proposed model because the previous models only learn part of the decisions and the rest is dealt with by traditional optimization techniques. Additionally, optimization methods like CVX are designed for one-step optimization, which are not for maximizing expected long-term rewards. In contrast, the proposed method learns all the actions by the DRL model to maximize expected long-term rewards.}

{\color{black}As shown in Fig.~\ref{fig:reward_cmp}, the proposed model is compared with the greedy algorithm and other DRL models. As we can see from the figure, the proposed DRL algorithm significantly outperforms the other models. In addition, the DRL models can get more rewards when the number of users increase because DRL models can learn better from more data, while the greedy algorithm is overwhelmed with more users and performs poorly. Moreover, the proposed model adopting end-to-end models has more freedom to choose the actions than existing DRL.} 
\begin{figure}[h!]
  \centering
  \begin{minipage}[b]{0.4\textwidth}
    \centering
    \includegraphics[width=3.0in]{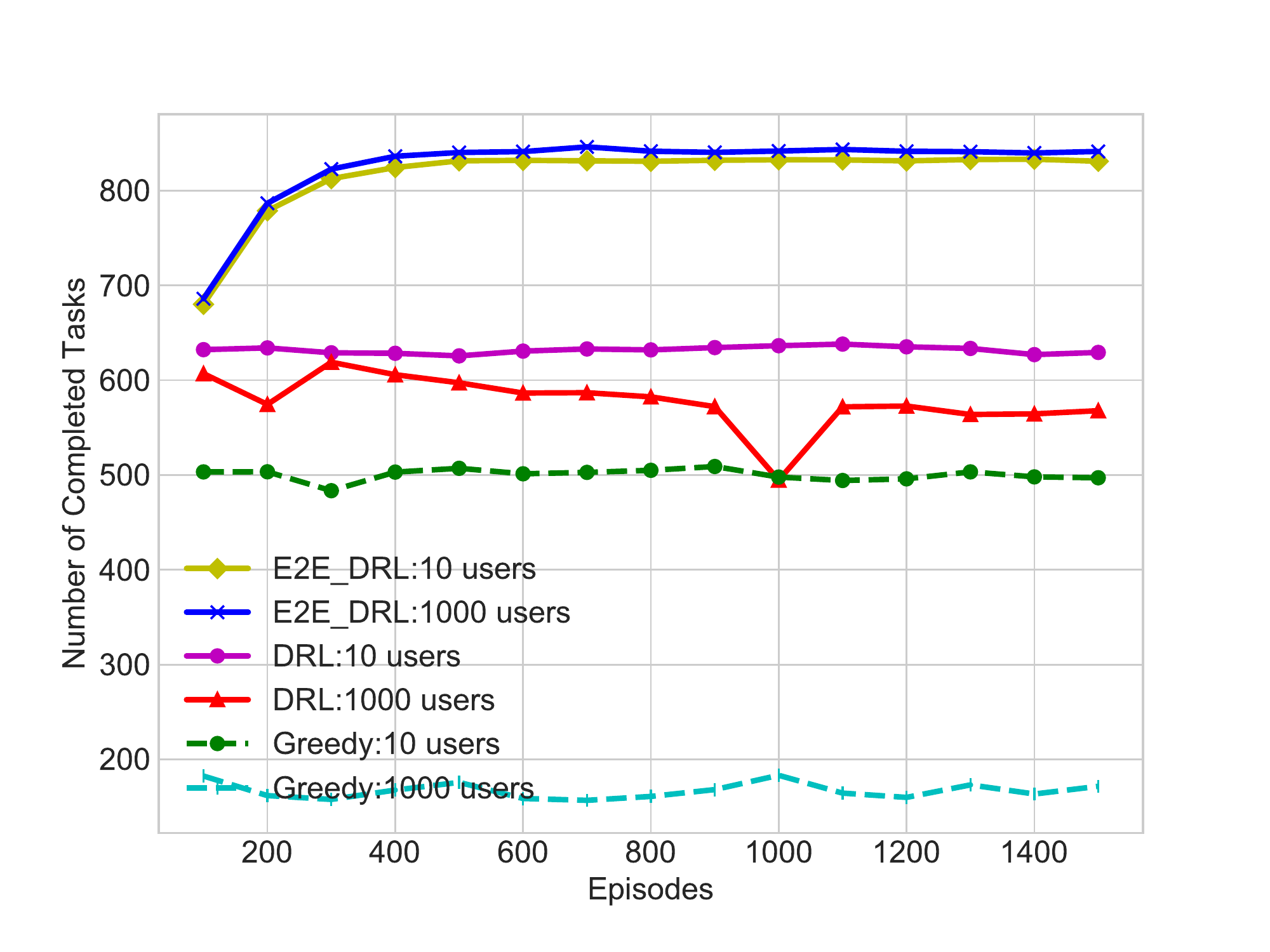}
    \caption{Tasks completion comparison}
    \label{fig:task_cmp}
  \end{minipage}
  \hfill
  \begin{minipage}[b]{0.4\textwidth}
    \centering
    \centering
    \includegraphics[width=3.0in]{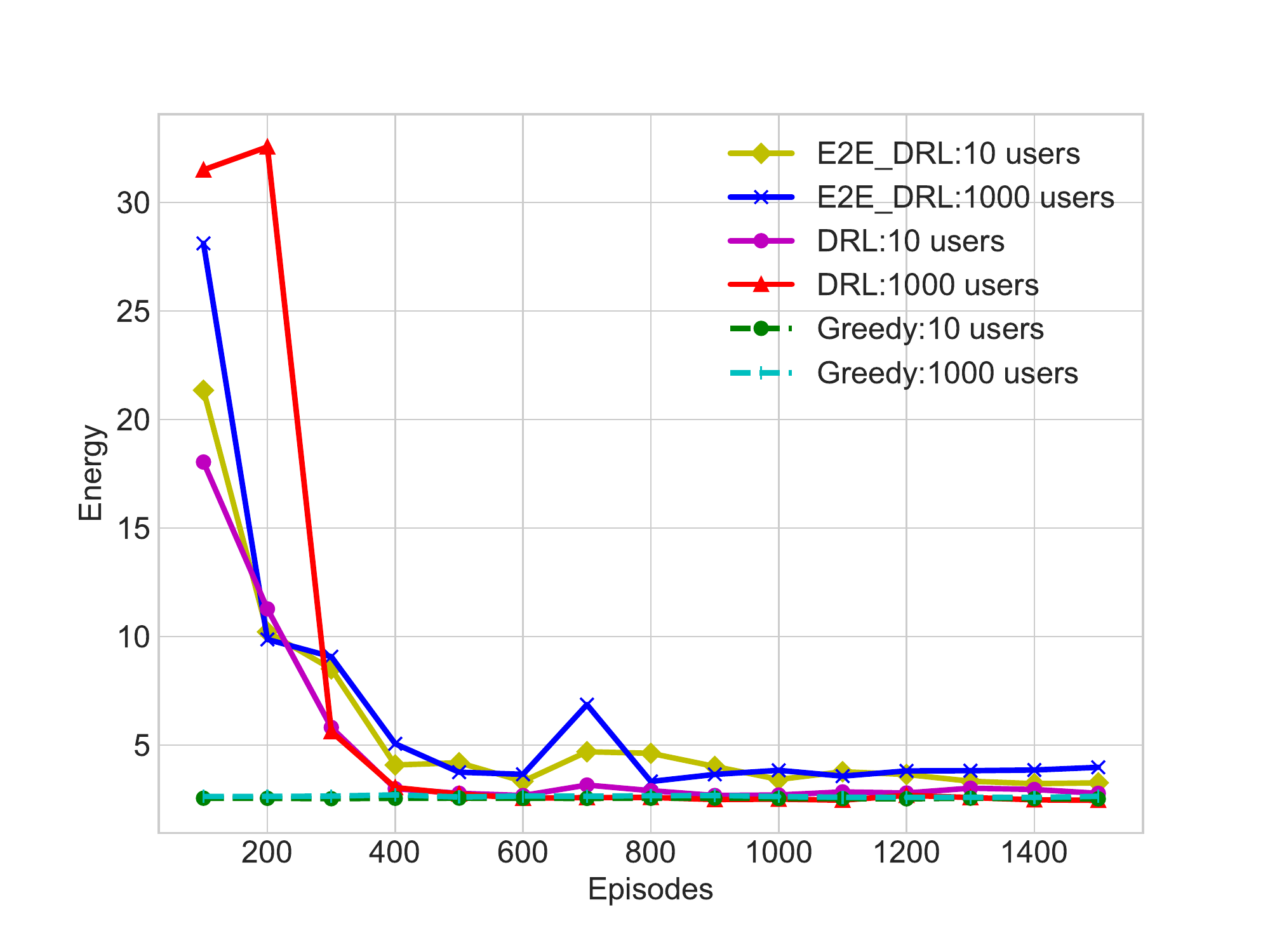}
    \caption{Average Energy Consumption}
    \label{fig:engergy_cmp}
  \end{minipage}
\end{figure}


{\color{black}For further comparison and analysis, the proposed model is compared with the existing DRL models and greedy algorithm in terms of the number of tasks completed and energy consumption, as shown in  Fig.~\ref{fig:task_cmp} and Fig.~\ref{fig:engergy_cmp}, respectively. Similar to the previous results for rewards, the proposed method outperforms the existing DRL models and the greedy algorithm. The performance of the proposed DRL models increases until they are converged to the optimal policies because the DRL models keep learning from the historical data stored in the replay buffer. On the contrary, the greedy algorithm remains almost the same. Similarly, the DRL models consume more energy in the beginning episodes because the model takes action with $\epsilon-greedy$ and randomly initialized parameters in the early episodes. The model can save more energy as the model learns from the acquired data and decreases random actions. Although the proposed DRL model is designed to maximize the long-term accumulated rewards, it also learns to reduce the energy cost over the time steps. The DRL model with CVX can significantly save energy consumption but does not increase the number of tasks because the DRL model can only control part of the decision variables. The computational cost is growing exponentially as the number of parameters and their search range increases.


\begin{figure}[h!]
  \centering
  \begin{minipage}[b]{0.4\textwidth}
    \centering
    \includegraphics[width=3.0in]{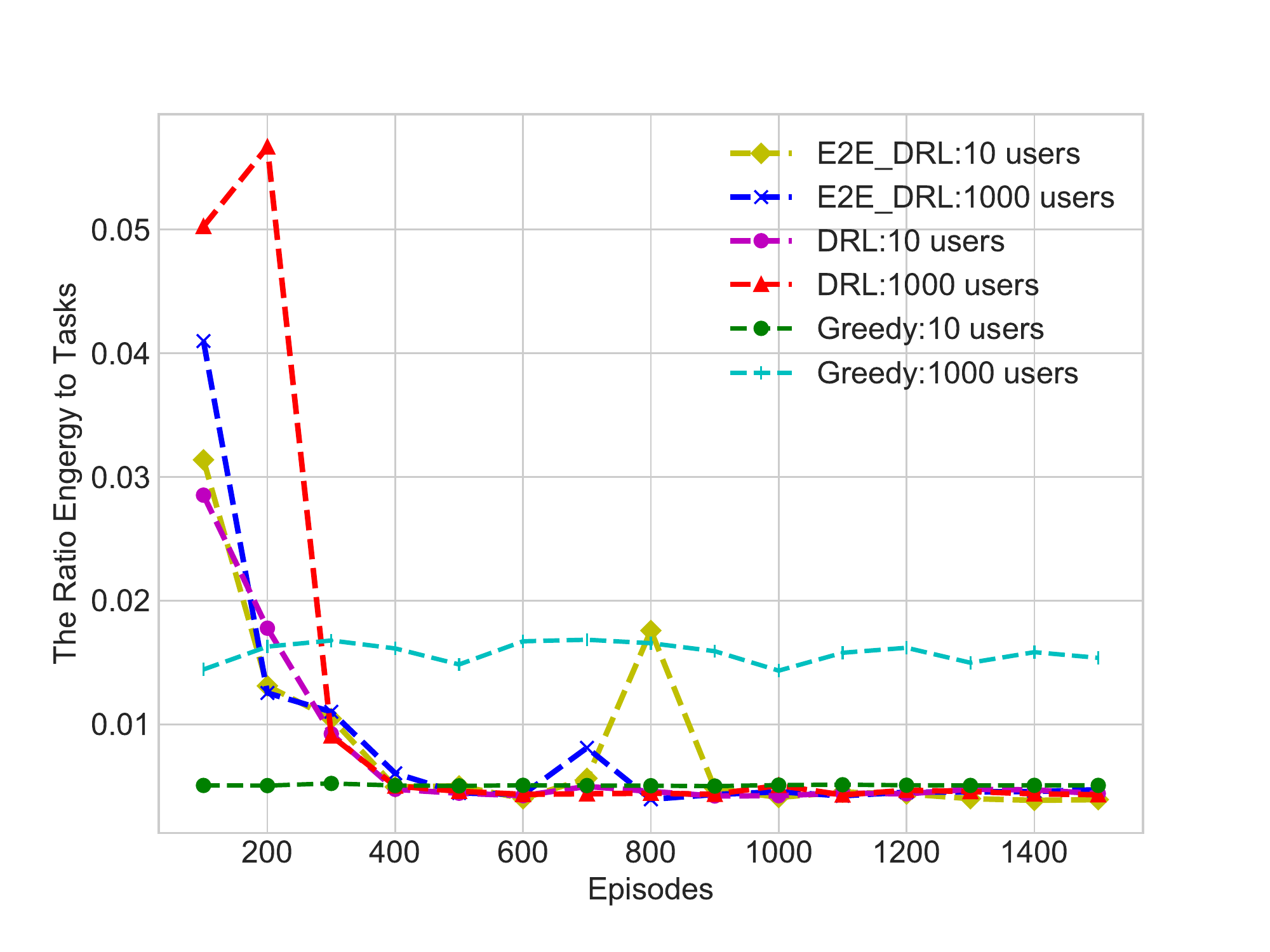}
    \caption{The Ratio of Energy To Tasks}
    \label{fig:engergy_to_tas_cmp}
  \end{minipage}
  \hfill
  \begin{minipage}[b]{0.4\textwidth}
    \centering
    \includegraphics[width=3.5in]{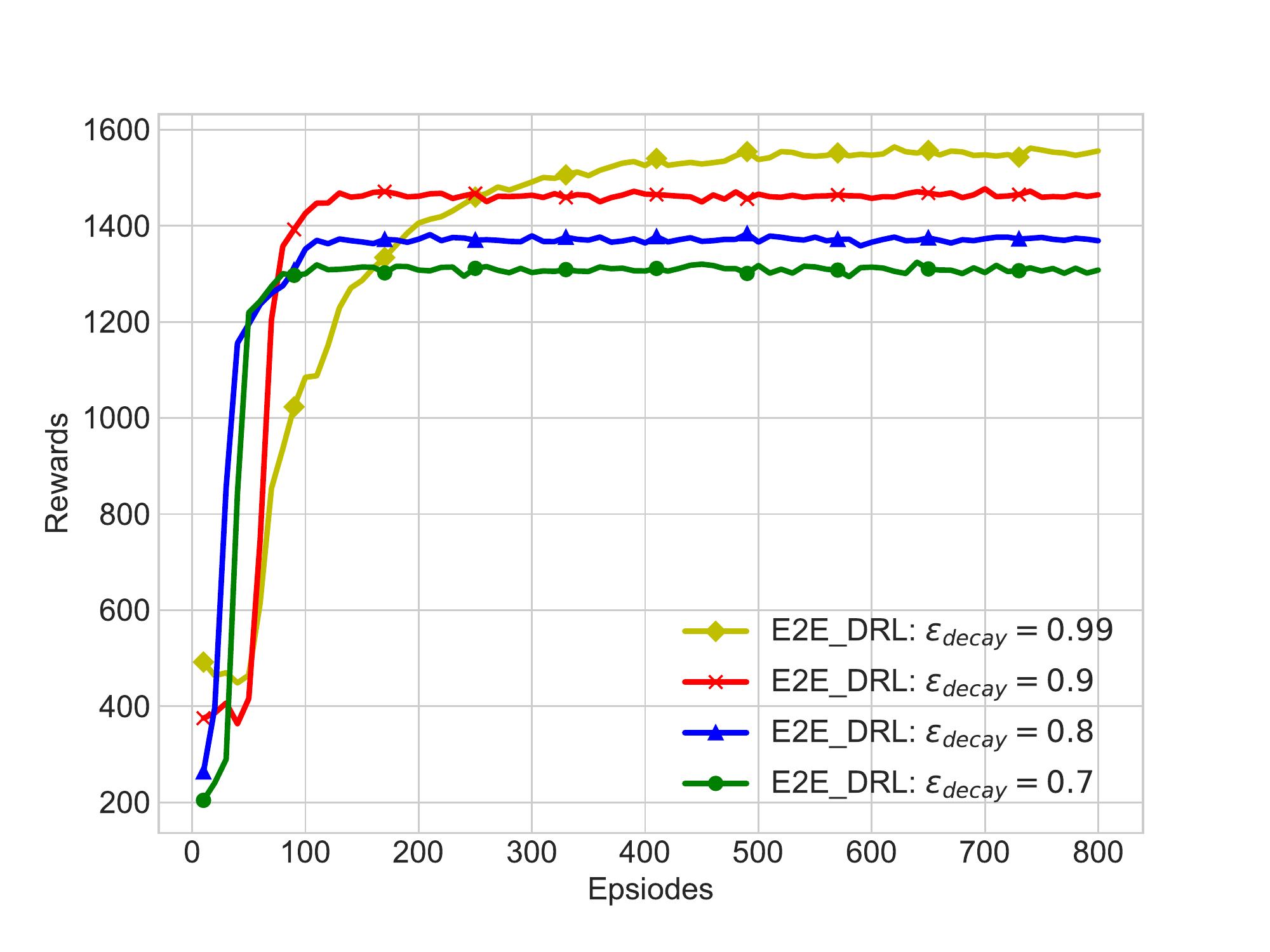}
    \caption{$\epsilon_{decay}$ for End-to-End DRL Model.}
    \label{fig:epsilong_cmp}
  \end{minipage}
\end{figure}

In the proposed DRL model, the network operator can balance the number of tasks completed and the energy cost by adjusting the weights in the reward function. Note that the randomness of the simulation caused the sharp peak at about from the $600^{th}$ to $800^{th}$ episode in Fig.~\ref{fig:engergy_cmp} and Fig.~\ref{fig:engergy_to_tas_cmp}.
The reasons for that are as follows. The energy consumption is proportional to the square of running CPU frequencies. Moreover, the computational cycles of the tasks in the simulator could range from cycles $8 \times 10^6$ to $1 \times 10^7$.  The learning algorithm incorporated $\epsilon$-greedy, which means the algorithm would take some random actions to explore the environment, and those actions sometimes significantly expend excessive energy. Therefore, the sharp point is raised by increasing computational cycles of tasks generated or the cost of exploitation.

Furthermore, Fig.~\ref{fig:epsilong_cmp} shows a comparison in the rewards and convergence with different $\epsilon_{decay}$ values. The proposed model uses $\epsilon_{decay}$ values to balance exploitation and exploration. The DRL model takes random actions with probability $\epsilon$ to explore, and $\epsilon=1$ at the beginning of learning. We want the model to take more random actions to explore the environment in the beginning and exploit the learned knowledge in the later episodes. Therefore, the value of $\epsilon$ is updated in each episode to decrease the exploration and increase exploitation, and the update operation is $\epsilon \leftarrow \epsilon_{decay} * \epsilon$. As we can see from Fig.~\ref{fig:epsilong_cmp}, the smaller $\epsilon_{decay}$ value, the faster the DRL model converges to its optimal policies. However, the models with small $\epsilon_{decay}$ may not find the optimal global policies if they stop to explore too early. Therefore, the DRL models large $\epsilon_{decay}$ values can achieve the models with small $\epsilon_{decay}$ values.
}


\section{conclusion}
{\color{black}
In this work, we have investigated the computation offloading problem in a dynamic MEC network. We propose an end-to-end DRL method to jointly optimize the edge server selection for offloading and the computing power allocation, with the objective of maximizing the number of tasks completed on time and minimizing energy consumption simultaneously. The proposed method can maximize long-term accumulated rewards instead of a one-time step. Moreover, the proposed model can make all the decisions without relying on other optimization functions to achieve joint optimization purposes. Applying the experience buffer replay and clip techniques to facilitate the DRL model training process can prevent the model from suffering from oscillation and divergence. Finally, simulation results are provided to demonstrate the effectiveness of the proposed method. For future work, we will study task partitioning in dynamic offloading, where tasks can be arbitrarily partitioned and then offloaded to edge servers.}

\nocite{*}
\bibliographystyle{IEEEannot}
\bibliography{annot}

\begin{thebibliography}{10}
\providecommand{\url}[1]{#1}
\csname url@rmstyle\endcsname
\providecommand{\newblock}{\relax}
\providecommand{\bibinfo}[2]{#2}
\providecommand\BIBentrySTDinterwordspacing{\spaceskip=0pt\relax}
\providecommand\BIBentryALTinterwordstretchfactor{4}
\providecommand\BIBentryALTinterwordspacing{\spaceskip=\fontdimen2\font plus
\BIBentryALTinterwordstretchfactor\fontdimen3\font minus
  \fontdimen4\font\relax}
\providecommand\BIBforeignlanguage[2]{{%
\expandafter\ifx\csname l@#1\endcsname\relax
\typeout{** WARNING: IEEEtran.bst: No hyphenation pattern has been}%
\typeout{** loaded for the language `#1'. Using the pattern for}%
\typeout{** the default language instead.}%
\else
\language=\csname l@#1\endcsname
\fi
#2}}

\bibitem{iots}
A.~Al-Fuqaha, M.~Guizani, M.~Mohammadi, M.~Aledhari, and M.~Ayyash, ``Internet
  of things: A survey on enabling technologies, protocols, and applications,''
  \emph{IEEE Communications Surveys \& Tutorials}, vol.~17, no.~4, pp.
  2347--2376, 2015.


\bibitem{MEC1}
Y.~Chen, N.~Zhang, Y.~Zhang, and X.~Chen, ``Dynamic computation offloading in
  edge computing for internet of things,'' \emph{IEEE Internet of Things
  Journal}, vol.~6, no.~3, pp. 4242--4251, 2018.


\bibitem{asheralieva2020bayesian}
A.~Asheralieva and D.~Niyato, ``Bayesian reinforcement learning and bayesian
  deep learning for blockchains with mobile edge computing,'' \emph{IEEE
  Transactions on Cognitive Communications and Networking}, 2020.


\bibitem{7120046}
N.~{Zhang}, N.~{Cheng}, A.~T. {Gamage}, K.~{Zhang}, J.~W. {Mark}, and
  X.~{Shen}, ``Cloud assisted hetnets toward 5g wireless networks,'' \emph{IEEE
  Communications Magazine}, vol.~53, no.~6, pp. 59--65, 2015.


\bibitem{MEC0}
T.~Q. {Dinh}, J.~{Tang}, Q.~D. {La}, and T.~Q.~S. {Quek}, ``Offloading in
  mobile edge computing: Task allocation and computational frequency scaling,''
  \emph{IEEE Transactions on Communications}, vol.~65, no.~8, pp. 3571--3584,
  Aug 2017.


\bibitem{op0_survey}
P.~{Mach} and Z.~{Becvar}, ``Mobile edge computing: A survey on architecture
  and computation offloading,'' \emph{IEEE Communications Surveys Tutorials},
  vol.~19, no.~3, pp. 1628--1656, thirdquarter 2017.


\bibitem{op1}
M.~{Chen} and Y.~{Hao}, ``Task offloading for mobile edge computing in software
  defined ultra-dense network,'' \emph{IEEE Journal on Selected Areas in
  Communications}, vol.~36, no.~3, pp. 587--597, March 2018.


\bibitem{DNN}
\BIBentryALTinterwordspacing
Y.~LeCun, Y.~Bengio, and G.~Hinton, ``Deep learning,'' \emph{Nature}, vol. 521,
  no. 7553, pp. 436--444, 2015. [Online]. Available:
  \url{https://doi.org/10.1038/nature14539}
\BIBentrySTDinterwordspacing


\bibitem{DNN2}
J.~Schmidhuber, ``Deep learning in neural networks: An overview,'' \emph{Neural
  Networks}, vol.~61, pp. 85 -- 117, 2015.


\bibitem{RNN}
L.~{Ale}, N.~{Zhang}, H.~{Wu}, D.~{Chen}, and T.~{Han}, ``Online proactive
  caching in mobile edge computing using bidirectional deep recurrent neural
  network,'' \emph{IEEE Internet of Things Journal}, vol.~6, no.~3, pp.
  5520--5530, June 2019.


\bibitem{RL}
R.~S. Sutton and A.~G. Barto, \emph{Reinforcement Learning: An Introduction},
  2nd~ed.\hskip 1em plus 0.5em minus 0.4em\relax USA: The MIT Press, 2018.


\bibitem{Q_learning}
S.~{Wang}, Y.~{Guo}, N.~{Zhang}, P.~{Yang}, A.~{Zhou}, and X.~S. {Shen},
  ``Delay-aware microservice coordination in mobile edge computing: A
  reinforcement learning approach,'' \emph{IEEE Transactions on Mobile
  Computing}, pp. 1--1, 2019.


\bibitem{op2_Q}
B.~{Dab}, N.~{Aitsaadi}, and R.~{Langar}, ``Q-learning algorithm for joint
  computation offloading and resource allocation in edge cloud,'' in \emph{2019
  IFIP/IEEE Symposium on Integrated Network and Service Management (IM)}, April
  2019, pp. 45--52.


\bibitem{q_Spectrum}
Z.~{Su}, M.~{Dai}, Q.~{Xu}, R.~{Li}, and S.~{Fu}, ``Q-learning-based spectrum
  access for content delivery in mobile networks,'' \emph{IEEE Transactions on
  Cognitive Communications and Networking}, vol.~6, no.~1, pp. 35--47, 2020.


\bibitem{DRL}
M.~Riedmiller, ``Neural fitted q iteration -- first experiences with a data
  efficient neural reinforcement learning method,'' in \emph{Machine Learning:
  ECML 2005}, J.~Gama, R.~Camacho, P.~B. Brazdil, A.~M. Jorge, and L.~Torgo,
  Eds.\hskip 1em plus 0.5em minus 0.4em\relax Berlin, Heidelberg: Springer
  Berlin Heidelberg, 2005, pp. 317--328.


\bibitem{drl_humanlevel}
V.~Mnih, K.~Kavukcuoglu, D.~Silver, A.~A. Rusu, J.~Veness, M.~G. Bellemare,
  A.~Graves, M.~Riedmiller, A.~K. Fidjeland, G.~Ostrovski, S.~Petersen,
  C.~Beattie, A.~Sadik, I.~Antonoglou, H.~King, D.~Kumaran, D.~Wierstra,
  S.~Legg, and D.~Hassabis, ``Human-level control through deep reinforcement
  learning,'' \emph{Nature}, vol. 518, no. 7540, pp. 529--533, Feb. 2015.


\bibitem{drl_vecular}
Z.~{Ning}, P.~{Dong}, X.~{Wang}, L.~{Guo}, J.~J. P.~C. {Rodrigues}, X.~{Kong},
  J.~{Huang}, and R.~Y.~K. {Kwok}, ``Deep reinforcement learning for
  intelligent internet of vehicles: An energy-efficient computational
  offloading scheme,'' \emph{IEEE Transactions on Cognitive Communications and
  Networking}, vol.~5, no.~4, pp. 1060--1072, 2019.


\bibitem{Li8906135}
J.~{Li}, X.~{Zhang}, J.~{Zhang}, J.~{Wu}, Q.~{Sun}, and Y.~{Xie}, ``Deep
  reinforcement learning-based mobility-aware robust proactive resource
  allocation in heterogeneous networks,'' \emph{IEEE Transactions on Cognitive
  Communications and Networking}, vol.~6, no.~1, pp. 408--421, 2020.


\bibitem{Wang8303773}
S.~{Wang}, H.~{Liu}, P.~H. {Gomes}, and B.~{Krishnamachari}, ``Deep
  reinforcement learning for dynamic multichannel access in wireless
  networks,'' \emph{IEEE Transactions on Cognitive Communications and
  Networking}, vol.~4, no.~2, pp. 257--265, 2018.


\bibitem{DRL_MEC1}
J.~{Chen}, S.~{Chen}, Q.~{Wang}, B.~{Cao}, G.~{Feng}, and J.~{Hu}, ``iraf: A
  deep reinforcement learning approach for collaborative mobile edge computing
  iot networks,'' \emph{IEEE Internet of Things Journal}, vol.~6, no.~4, pp.
  7011--7024, Aug 2019.


\bibitem{DRL_MEC2}
L.~{Huang}, S.~{Bi}, and Y.~J. {Zhang}, ``Deep reinforcement learning for
  online computation offloading in wireless powered mobile-edge computing
  networks,'' \emph{IEEE Transactions on Mobile Computing}, pp. 1--1, 2019.


\bibitem{9085261}
J.~{Chen}, Z.~{Wei}, S.~{Li}, and B.~{Cao}, ``Artificial intelligence aided
  joint bit rate selection and radio resource allocation for adaptive video
  streaming over f-rans,'' \emph{IEEE Wireless Communications}, vol.~27, no.~2,
  pp. 36--43, 2020.


\bibitem{9120235}
J.~{Du}, F.~R. {Yu}, G.~{Lu}, J.~{Wang}, J.~{Jiang}, and X.~{Chu},
  ``Mec-assisted immersive vr video streaming over terahertz wireless networks:
  A deep reinforcement learning approach,'' \emph{IEEE Internet of Things
  Journal}, pp. 1--1, 2020.


\bibitem{SchulmanWDRK17}
\BIBentryALTinterwordspacing
J.~Schulman, F.~Wolski, P.~Dhariwal, A.~Radford, and O.~Klimov, ``Proximal
  policy optimization algorithms,'' \emph{CoRR}, vol. abs/1707.06347, 2017.
  [Online]. Available: \url{http://arxiv.org/abs/1707.06347}
\BIBentrySTDinterwordspacing


\bibitem{cycles}
J.~{Zhang}, X.~{Hu}, Z.~{Ning}, E.~C.~. {Ngai}, L.~{Zhou}, J.~{Wei},
  J.~{Cheng}, and B.~{Hu}, ``Energy-latency tradeoff for energy-aware
  offloading in mobile edge computing networks,'' \emph{IEEE Internet of Things
  Journal}, vol.~5, no.~4, pp. 2633--2645, Aug 2018.


\bibitem{channel}
R.~{Bultitude}, ``Measurement, characterization and modeling of indoor 800/900
  mhz radio channels for digital communications,'' \emph{IEEE Communications
  Magazine}, vol.~25, no.~6, pp. 5--12, June 1987.


\bibitem{8016573}
Y.~{Mao}, C.~{You}, J.~{Zhang}, K.~{Huang}, and K.~B. {Letaief}, ``A survey on
  mobile edge computing: The communication perspective,'' \emph{IEEE
  Communications Surveys Tutorials}, vol.~19, no.~4, pp. 2322--2358, 2017.


\bibitem{7542156}
Y.~{Wang}, M.~{Sheng}, X.~{Wang}, L.~{Wang}, and J.~{Li}, ``Mobile-edge
  computing: Partial computation offloading using dynamic voltage scaling,''
  \emph{IEEE Transactions on Communications}, vol.~64, no.~10, pp. 4268--4282,
  2016.


\bibitem{8352664}
S.~{Guo}, J.~{Liu}, Y.~{Yang}, B.~{Xiao}, and Z.~{Li}, ``Energy-efficient
  dynamic computation offloading and cooperative task scheduling in mobile
  cloud computing,'' \emph{IEEE Transactions on Mobile Computing}, vol.~18,
  no.~2, pp. 319--333, 2019.


\bibitem{normalize}
G.~H. Golub and C.~F.~V. Loan, \emph{Matrix Computation}, 4th~ed.\hskip 1em
  plus 0.5em minus 0.4em\relax Baltimore: The Johns Hopkins University Presss,
  2013, p.~71.


\bibitem{adam}
D.~P. Kingma and J.~Ba, ``Adam: {A} method for stochastic optimization,''
  \emph{CoRR}, vol. abs/1412.6980, 2014.


\bibitem{MEC_challenges}
Y.~{Yu}, ``Mobile edge computing towards 5g: Vision, recent progress, and open
  challenges,'' \emph{China Communications}, vol.~13, no. Supplement2, pp.
  89--99, N 2016.


\end{thebibliography}

\end{document}